\documentclass[aps,prl,showpacs,twocolumn,amsmath,amssymb]{revtex4-1}
\usepackage{graphicx}
\usepackage{dcolumn}
\usepackage{bm}
\usepackage{units}
\usepackage{upgreek}
\usepackage{ucs}
\usepackage{color}
\usepackage{layouts}
\usepackage{hyperref}
\usepackage[all]{hypcap}
\usepackage{standalone}
\usepackage{tikz}
\preprint{APS/123-QED}

\begin{document}

\title{Giant cross section for molecular ion formation in ultracold Rydberg gases}
\author{Thomas Niederpr\"um$^1$}
\author{Oliver Thomas$^{1,2}$}
\author{Torsten Manthey$^1$}
\author{Tobias M. Weber$^1$}
\author{Herwig Ott$^1$}
\email{ott@physik.uni-kl.de}
\affiliation{$^1$Research Center OPTIMAS, Technische Universit\"at Kaiserslautern, 67663 Kaiserslautern, Germany}
\affiliation{$^2$Graduate School Materials Science in Mainz, Staudinger Weg 9, 55128 Mainz, Germany}

\begin{abstract}
We have studied the associative ionization of a Rydberg atom and a ground state atom in an ultracold Rydberg gas. The measured scattering cross section is three orders of magnitude larger than the geometrical size of the produced molecule. This giant enhancement of the reaction kinetics is due to an efficient directed mass transport which is mediated by the Rydberg electron. We also find that the total inelastic scattering cross section is given by the geometrical size of the Rydberg electron's wavefunction.
\end{abstract}

\maketitle



Molecule formation in dilute gases or plasmas usually happens via two-body collisions and is relevant, e.g., for the production of molecules in interstellar space  \cite{Geballe1996} or during the early stage of the universe \cite{Dalgarno2005}. One prominent class of such collisions is the associative ionization between a ground state atom and a (highly) excited atom. The latter can result from electron capture in a plasma or photon absorption. The formation of a bound molecule requires the release of binding energy which is realized by the ejection of an electron \cite{Mihajlov2012}. Associative ionization involving Rydberg atoms has been studied in detail in hot atomic beam experiments and cross sections on the order of the geometrical size of the formed molecules have been found \cite{Klucharev1980, Cheret1982, Barbier1987}.

Ultracold Rydberg gases allow to extend the study of this fundamental chemical reaction to the low-energy limit, where a new interaction mechanism \cite{Fermi1934} has drawn researchers' attention in the last years: The scattering between the electron in a Rydberg state and a ground state atom creates a potential for the atom that reaches far from the Rydberg atoms core. This potential gives rise to ultra long range Rydberg molecules \cite{Bendkowsky2009}, trilobite molecules \cite{Booth2014} and was found to induce phonons inside of a Bose-Einstein-Condensate \cite{Balewski2013}. The understanding of this interaction and it's role in associative ionization is also important to fully exploit the potential of ultracold Rydberg systems to study many-body quantum phenomena \cite{Schausz2012,Schempp2014,Weber2015} beyond the frozen gas approximation.

Here, we study the associative ionization of a rubidium atom in a Rydberg p-state (principal quantum number $n=30 - 60$) and a ground state atom at ultracold temperatures. The measured scattering cross section is three orders of magnitude larger than the geometrical size of the produced molecular ion. We attribute this enhancement to a directed mass transport of the ground state atom towards the ionic core of the Rydberg atom. This transport mechanism is mediated by the scattering between the Rydberg electron and the ground state atom. The formation of the molecular ion happens, when the two collision partners are close enough that the released binding energy suffices to eject the excited electron via resonant dipole interaction \cite{Mihajlov2012}. The appearance of a highly efficient mass transport which accelerates the reaction kinetics, the quasi-free nature of the Rydberg electron and the fact that the electron mediates the reaction but is not part of the reaction product allow us to draw a connection to a generic catalytic reaction, where the electron acts as a catalyst \cite{Studer2014}. We also find, that the total inelastic cross section between a Rydberg atom and a ground state atom is as large as the size of the Rydberg electron wave function. In turn, the probability for an elastic collision between a Rydberg atom and a ground state atom is practically negligible for the investigated parameter range.


\begin{figure}
\begin{center}
\includegraphics[width=\columnwidth]{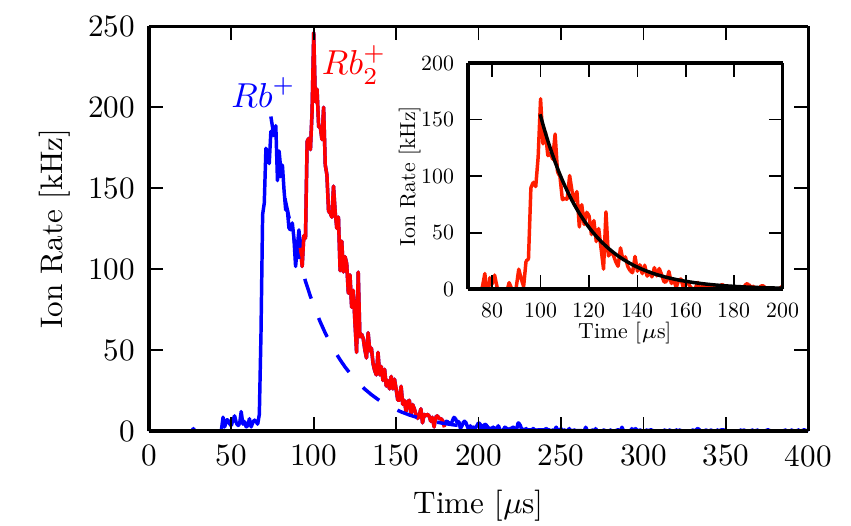}
\end{center}
\caption{(Color online). Time of flight spectrum for the $30P_{3/2}$~-~state at a peak density of $n_0 = 3 \times 10^{13}\,\mathrm{cm}^{-3}$ showing the $Rb^+$- and $Rb_2^+$~-~ peaks. After subtracting the fitted decay of the $Rb^+$~-~peak, the signal of the molecular ions alone is obtained (inset).} 
\label{fig:tofspectrum}
\end{figure}

The experimental apparatus is described in detail in Ref.\,\cite{Manthey2014}. In brief, we produce a thermal cloud of $N = 4 \times 10^5$ rubidium atoms in the $F=1,m_F=+1$ ground state at a temperature of $T = 200\,\mathrm{nK}$ in a crossed dipole trap ($\lambda = 1064\,\mathrm{nm}$) with trapping frequencies of $\omega_\mathrm{x,y,z} = 2\pi \times (77, 77, 94)\,\mathrm{Hz}$. In order to prepare clouds of different densities the atom number is reduced by release and recapture of the atoms. This way, the density of the cloud can be changed by two orders of magnitude while the temperature increases at most by a factor of two and the trap geometry is constant for all measurements. 
A single-photon excitation to Rydberg $nP$ - states is driven by a frequency doubled dye-laser at a wavelength of $\lambda = 297\,\mathrm{nm}$. By applying an offset magnetic field of $35\,\mathrm{Gs}$ we are able to individually address the Zeeman substates of the Rydberg levels. For the measurements presented here we are always addressing the $m_j = +3/2$ - state. A pulsed excitation scheme with a width of $1\,\mu\mathrm{s}$  and a pulse spacing of $1\,\mathrm{ms}$ is repeated as long as the loss of atoms from the sample due to excitation and ionization does not exceed 20\%. After the excitation pulse, the Rydberg excitations are converted into ions by photoionization through the trapping lasers (total intensity $I=7.2$\,kW/cm$^2$), black body radiation as well as by associative ionization. The ions are extracted by a small electric field of $80\,\mathrm{mV/cm}$. A typical time-of-flight (TOF) spectrum is shown in Fig.~\ref{fig:tofspectrum}. The two distinct peaks in the resulting spectrum can be unambiguously attributed to $Rb^+$ and $Rb_2^+$ ions.

The $Rb^+$ ions originate from photoionization by the trapping lasers and black body radiation with rate $\gamma_\mathrm{PI} = \gamma^\mathrm{YAG}_\mathrm{PI} + \gamma^\mathrm{BB}_\mathrm{PI}$. A dedicated measurement in very dilute clouds was performed in order to determine the photoionization cross section of the trapping light (Tab.~\ref{tab:crosssections}). Due to the low intensity of the trapping laser in the presented measurements the photoionization rate $\gamma^\mathrm{YAG}_\mathrm{PI} = \sigma_\mathrm{PI}^\mathrm{YAG}\frac{I}{h\nu} < 500\,\mathrm{Hz}$  is negligible compared to the natural decay of the Rydberg states involved. Also the ionization by black-body photons $\gamma^\mathrm{BB}_\mathrm{PI}$ does not exceed $300\,\mathrm{Hz}$ for our experimental settings \cite{Beterov2007}. Therefore the decay of the $Rb^+$ - peak in the TOF measurement is solely determined by the natural decay $\gamma_\mathrm{n}$ of the Rydberg state. Fitting this decay by an exponential function and subtracting it from the data, the pure signal of the $Rb_2^+$~-~ions is obtained (inset of Fig.~\ref{fig:tofspectrum}). Such TOF measurements enable us to study systematically the two-body dynamics involved in the formation of $Rb_2^+$ molecular ions. 


\begin{figure}
\begin{center}
\includegraphics[width=\columnwidth]{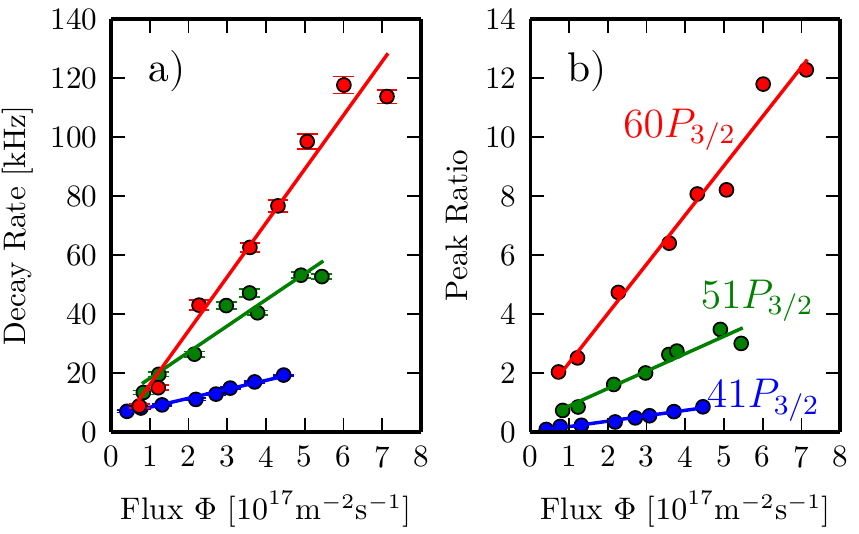}
\end{center}
\caption{(Color online). (a) The slope of the linear dependency between the decay rate of the $Rb_2^+$ - peak and the atom flux $\Phi$ through the Rydberg atom reveals the total inelastic scattering cross section $\sigma_\mathrm{inel}$ between a Rydberg and a ground state atom. (b) The scaling of the peak ratio $R_\mathrm{AI}/R_\mathrm{PI}$ reveals the partial scattering cross section for the creation of molecular ions $\sigma_\mathrm{AI}$.} 
\label{fig:scaling}
\end{figure}

To get a quantitative understanding of the molecular ion formation process we use the following rate model. We first assume the density of ground state atoms, $n(r) = n_0 e^{-r^2/2\sigma^2}$, to be isotropic with width $\sigma$. After the excitation pulse, the time evolution of the density of Rydberg atoms $n_\mathrm{Ry}(r, t)$ is determined by the differential equation
\begin{align}\label{eq:decay}
\dot{n}_\mathrm{Ry}(r, t) = -\left[\gamma_\mathrm{n} + \gamma_\mathrm{PI} + \gamma_\mathrm{coll}(r)\right]n_\mathrm{Ry}(r, t).
\end{align} 
Here $\gamma_\mathrm{coll}(r) = n(r)  \sigma_\mathrm{inel} \bar{v}_\mathrm{coll}$, is the total inelastic collision rate between ground state and Rydberg atoms and the average collision velocity $\bar{v}_\mathrm{coll} = 1.56  \sqrt{2} \bar{v}_\mathrm{th}$ is related to the average thermal velocity $\bar{v}_\mathrm{th}$. The numerical prefactor in the collision velocity takes into account the additional recoil momentum that is transferred to the Rydberg atom upon excitation. The collision rate $\gamma_\mathrm{coll}(r)$ accounts for any scattering process of a Rydberg atom with a ground state atom which destroys the Rydberg excitation. A fraction of these collisions lead to the formation of molecular ions for which we define the rate $\gamma_\mathrm{AI}(r) = n(r) \sigma_\mathrm{AI} \bar{v}_\mathrm{coll}$, where $\sigma_\mathrm{AI}$ denotes the partial scattering cross section for associative ionization. Eq.\,\ref{eq:decay} leads to an exponential decay at every point of the sample. Integration over the sample leads to
\begin{align}
\label{eq:rpi}
R_\mathrm{PI}(t) &= \gamma_\mathrm{PI} e^{-(\gamma_\mathrm{n} + \gamma_\mathrm{PI})t} \int n_\mathrm{Ry}(r)  e^{-\gamma_\mathrm{coll}(r)t}~\mathrm{d}^3r\\
\label{eq:rai}
R_\mathrm{AI}(t) &= e^{-(\gamma_\mathrm{n} + \gamma_\mathrm{PI})t} \int n_\mathrm{Ry}(r) \gamma_\mathrm{AI}(r) e^{-\gamma_\mathrm{coll}(r)t} ~\mathrm{d}^3r,
\end{align}
where $R_\mathrm{PI}(t)$ ($R_\mathrm{AI}(t)$) denotes the total rate for photoionization (associative ionization). Taking the decay rate of the latter one at $t=0$ yields:
\begin{align}
\label{eq:gamma0}
\gamma_\mathrm{coll}^0 =\left. \frac{\dot{R}_\mathrm{AI}}{R_\mathrm{AI}}\right|_{t = 0} = -\sqrt{\frac{8}{27}}\sigma_\mathrm{inel}\Phi - (\gamma_\mathrm{n} + \gamma_\mathrm{PI}).
\end{align}
where $\Phi=n_0\bar{v}_\mathrm{coll}$ is the atom flux impinging on the electron in the trap center. Here, we have additionally assumed that the system is not blockaded \cite{Urban2009, Gaetan2009} and the density of Rydberg atoms follows the density of ground state atoms, $n_\mathrm{Ry}(r) \propto n(r)$. This assumption is fulfilled for all data sets, except for those at the highest fluxes\footnote{The Rydberg blockade becomes relevant when the peak density of the cloud $n_0$ exceeds the volume of the blockade sphere $r_\mathrm{B}^3$ }.

\begin{figure}
\begin{center}
\includegraphics[width=\columnwidth]{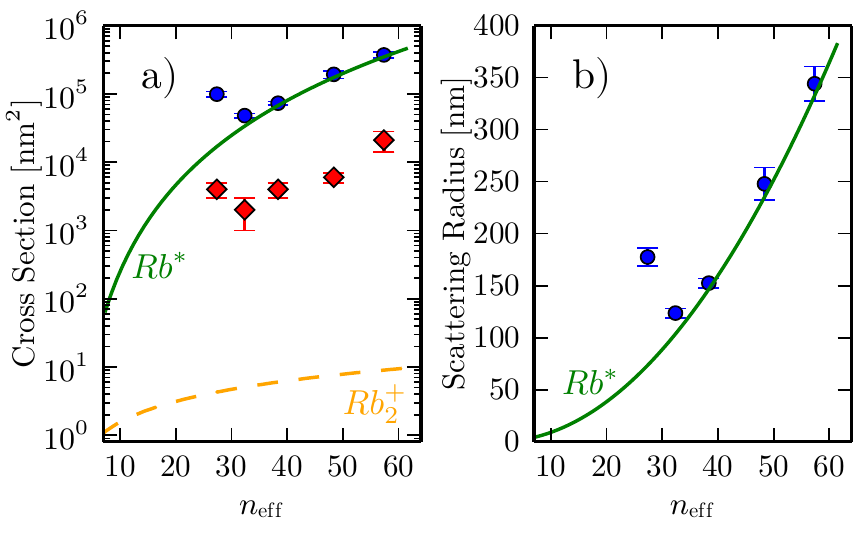}
\end{center}
\caption{(Color online). Scaling of the measured scattering properties with the effective principal quantum number $n_\mathrm{eff}$. (a) The cross section for associative ionization $\sigma_\mathrm{AI}$ (red diamonds) exceeds the geometrical cross section of the formed $Rb_2^+$ molecules (orange dashed line) by three orders of magnitude. The total inelastic scattering cross section $\sigma_\mathrm{inel}$ (blue dots) is in very good agreement with the geometrical cross section of the Rydberg atom (solid green line). (b) The characteristic scattering radius coincides with the size of the Rydberg atom, unveiling that inelastic collisions start to appear as soon as the ground state atom enters the electronic wavefunction of the Rydberg atom.} 
\label{fig:sizeCompare}
\end{figure}

We have taken a full set of TOF spectra for the Rydberg p-states $n=30,35,41,51,60$ and for varied ground state densities, covering almost two orders of magnitude in atomic flux. The results are summarized in Fig.~\ref{fig:scaling}. Comparing the experimental results to Eq.~\eqref{eq:gamma0}, the total inelastic scattering cross section $\sigma_\mathrm{inel}$ is obtained by fitting the initial decay of the molecular peak and extracting the slope of the obtained decay rates as a function of the atomic flux (Fig.~\ref{fig:scaling}~a). The results for $\sigma_\mathrm{inel}$ are given in Tab.~\ref{tab:crosssections}. These values are comparable to the geometrical cross section of the Rydberg atom. Indeed, we can calculate the characteristic scattering radius $l = \sqrt{\sigma_\mathrm{inel}/\pi}$ and compare it to the radial distance of the outermost maximum of the Rydberg electron wavefunction in Fig.~\ref{fig:sizeCompare}). Except for the $30P_{3/2}$ state, both values are in excellent agreement. Hence, every ground state atom which enters the Rydberg electron wave function, leads to a decay of the Rydberg excitation. It is therefore the size of the electron's wave function which sets the limit for the inelastic collisional cross section. In the case of the $30P_{3/2}$ state however this limit is exceeded by a factor of two, which we attribute to Rydberg~-~Rydberg collisions appearing at higher numbers of excitations due to the strong coupling to low lying states and the reduced blockade radius. Such collisions are not accounted for in the rate model.

Undergoing inelastic scattering the ground state atom can enter a molecular state with the ionic core of the Rydberg atom through associative ionization. This is revealed by the presence of molecular ions in the TOF spectrum. Analyzing the relative peak heights, the partial scattering cross section for associative ionization $\sigma_\mathrm{AI}$ can be obtained by taking the ratio of Eq.~\eqref{eq:rai} and Eq.~\eqref{eq:rpi} at time $t=0$:
\begin{align}
\label{eq:ratio}
\left. \frac{R_\mathrm{AI}}{R_\mathrm{PI}}\right|_{t = 0} = \frac{\sigma_\mathrm{AI}}{\sqrt{8}\gamma_\mathrm{PI}}\Phi.
\end{align}
To extract $\sigma_\mathrm{AI}$, we fit a linear slope to the peak ratio in dependence of the atomic flux (Fig.\,\ref{fig:scaling}b). The results are summarized in Tab.~\ref{tab:crosssections} and visualized in Fig.~\ref{fig:sizeCompare} a). It is remarkable that the cross section reaches values up to $2.1\times10^4\,\mathrm{nm}^2$. This is orders of magnitude larger than previously measured cross sections for thermal beams that were found to peak at $3.9\,\mathrm{nm}^2$ for $n=11$ \cite{Klucharev1980}, which agrees with the geometrical cross section of the formed molecule, estimated from the size given by the classical turning point of the $Rb_2^+$ molecular potential (Tab.~\ref{tab:crosssections}). Also, the cross section for associative ionization increases with the size of Rydberg atom - contrary to experiments in hot atomic beams, where the opposite dependence is observed \cite{Klucharev1980}. 

\begin{table}[t]
 \caption{Measured cross sections for photo ionization by the YAG trapping light $\sigma_\mathrm{PI}^\mathrm{YAG}$, for inelastic scattering between a Rydberg atom and a ground state atom $\sigma_\mathrm{inel}$ and for associative ionization $\sigma_\mathrm{AI}$. For comparison the theoretical geometrical cross section $\sigma^{\mathrm{Rb}_2^+}_\mathrm{geo}$ of the formed $\mathrm{Rb}_2^+$ molecules is presented.}

 \begin{tabular}{c|c|c|c|c}
 state & $\sigma_\mathrm{PI}^\mathrm{YAG}~[\mathrm{kb}]$ & $\sigma_\mathrm{inel}~[\mathrm{nm}^2]$ & $\sigma_\mathrm{AI}~[\mathrm{nm}^2]$ & $\sigma^{\mathrm{Rb}_2^+}_\mathrm{geo}~[\mathrm{nm}^2]$\\
 \hline
 $30P_{3/2}$ & 16(5) & $9.9(9)\times 10^4$ & $4(1)\times 10^3$ & 4.3\\
 $35P_{3/2}$ & 22(6) & $4.8(3)\times 10^4$ & $2(1)\times 10^3$ & 5.1\\
 $41P_{3/2}$ & 14(4) & $7.3(4)\times 10^4$ & $4(1)\times 10^3$ & 6.0\\
 $51P_{3/2}$ & 6(2) & $1.9(3)\times 10^5$ & $6(1)\times 10^3$ & 7.6\\
 $60P_{3/2}$ & 8(3) & $3.7(4)\times 10^5$ & $21(7)\times 10^3$ & 9.2\\
 \end{tabular}
 \label{tab:crosssections}
 \end{table}

\begin{figure}[b]
\begin{center}
\begin{tikzpicture}
	\node (plot) at (0,0) {\includegraphics[width=\columnwidth]{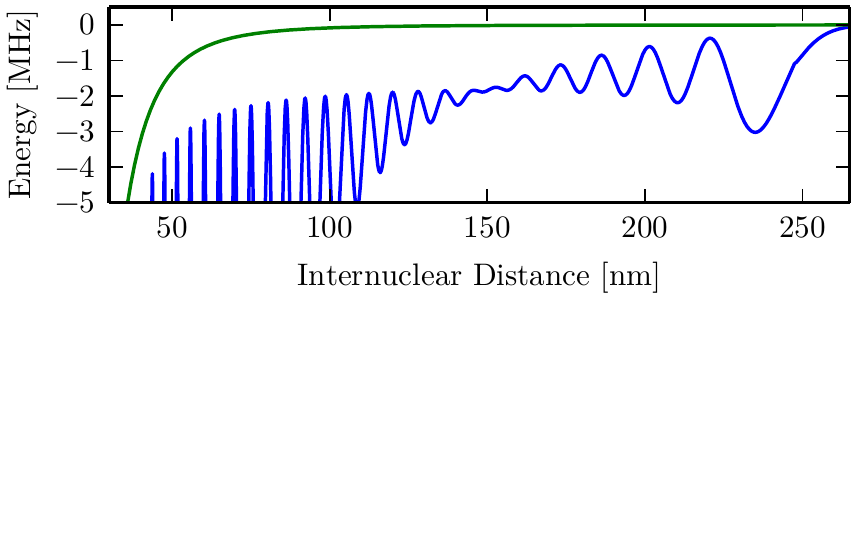}};
	\node at (-78pt,63pt) {\large a)};
	\node (img) at (-60pt,-42pt) {\includegraphics[width=0.4\columnwidth]{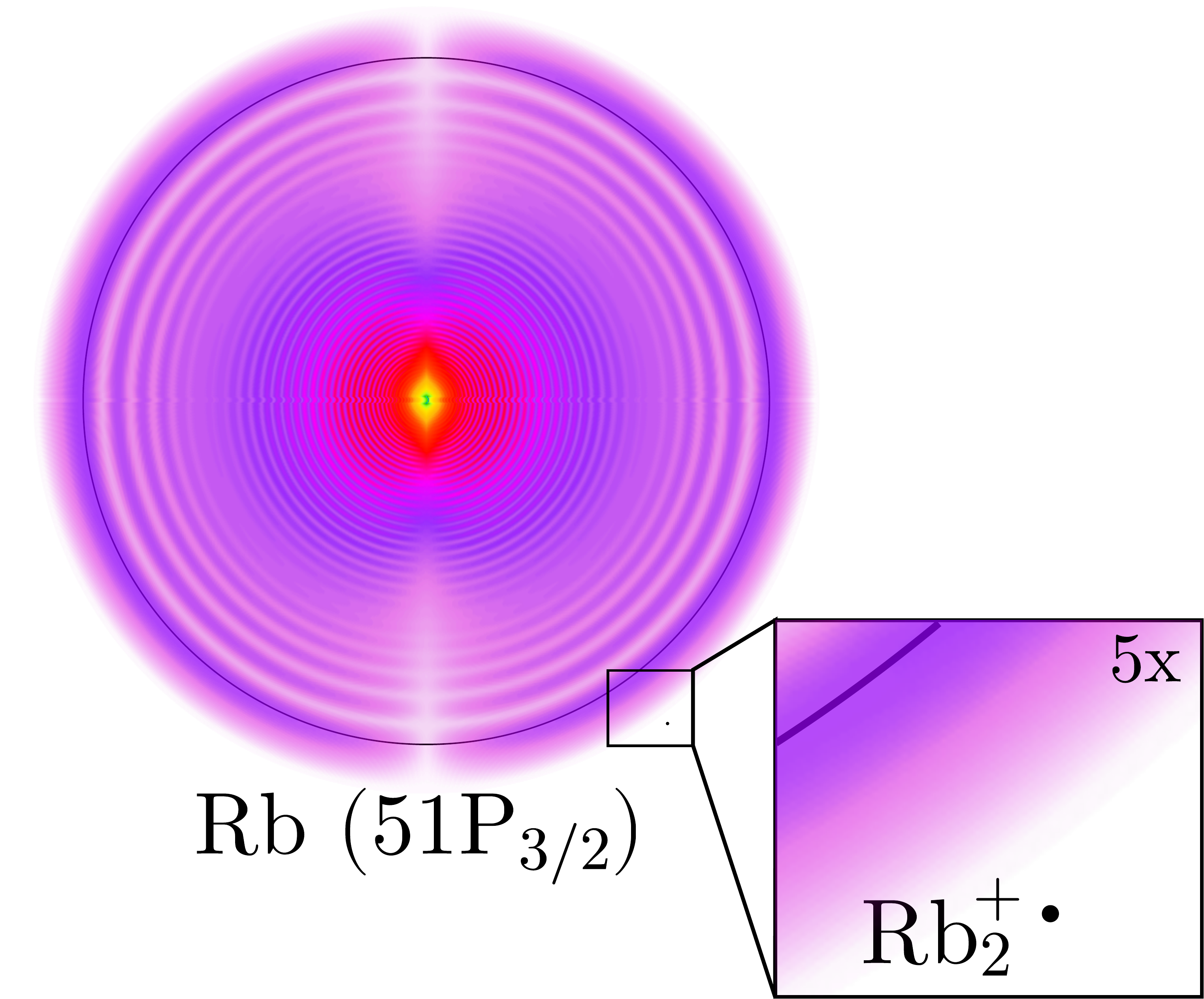}};	
	\node at (-115pt,-25pt) {\large b)};
	\node (graph) at (70pt,-50pt) {\includegraphics[width=0.43\columnwidth]{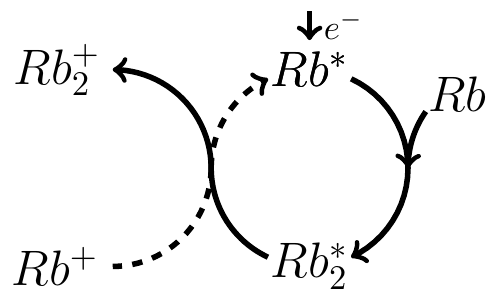}};
	\node at (15pt,-25pt) {\large c)};
\end{tikzpicture}
\end{center}
\caption{(Color online). (a) Potential for the ground state atom created by the interaction with the Rydberg core alone (green) and in combination with the  Rydberg electron (blue). The potential is calculated in perturbation theory for triplet scattering. (b) Comparison of the $51P_{3/2}$ electronic wavefunction with the size of the formed $Rb_2^+$-molecule. (c) Catalytic cycle enabled by the presence of the Rydberg electron. The solid line illustrates the quasi-catalytic effect of the electron observed in our experiments. The electron of the Rydberg atom $Rb^*$ vastly increases the probability of an inelastic collision with a ground state atom $Rb$, ultimately leading to a metastable excited $Rb_2^*$ molecule that subsequently ionizes into $Rb_2^+$ through the dipole resonant mechanism \cite{Mihajlov2012}. In an ultracold plasma the recombination of $Rb^+$ and $e^-$ into Rydberg states (dashed line) could complete the catalytic cycle and provide a very efficient mechanism for the creation of $Rb_2^+$ molecular ions.} 
\label{fig:potential}
\end{figure}
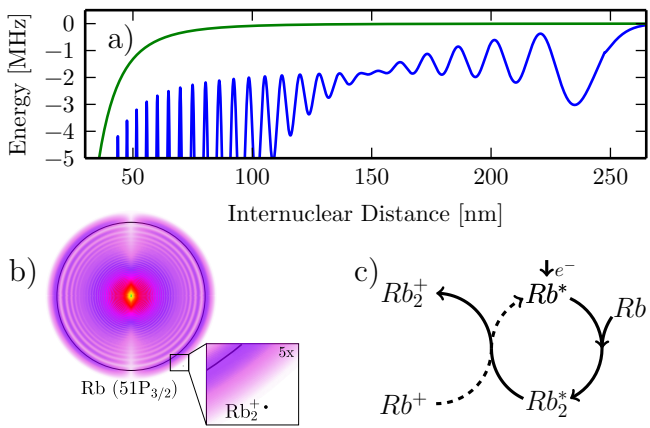

A microscopic explanation of our findings can be given, considering the Fermi contact interaction between the Rydberg electron and the ground state atom \cite{Greene2000}. The interplay between s- and p-wave scattering creates an oscillatory potential with an attractive envelope in the whole volume occupied by the electron. For small internuclear separations the polarization of the ground state atom with polarizability $\alpha_0$ by the ionic core of the Rydberg atom starts to dominate the interaction, leading to a $C_4/r^4$~-~type potential ($C_4 = \frac{\alpha_0 q^2}{32 \pi^2 \epsilon_0^2} = 5.5\times 10^{-57}\,\mathrm{Jm}^4$). Note that its influence is negligible at the outer parts of the electron's wave function. Fig.~\ref{fig:potential} shows a perturbation theory calculation of the triplet contribution to the interaction potential for a ground state atom inside the $51P_{3/2}$ electronic wavefunction combined with the $1/r^4$~-~type interaction. A full calculation of the triplet and singlet scattering potential taking into account the hyperfine interaction in the ground state atom is given in Ref.~\cite{Anderson2014}. The potential created by the electron allows in principle for a transport of the incoming particle towards the ionic core. Integrating the equation of motion of a classical particle in the potential shows that the transport can happen at a timescale of a few microseconds, much faster than the lifetime of the Rydberg atom. The real problem, of course, is much more complex, as the thermal deBroglie wavelength of the colliding atoms is of the same size as the Rydberg electron's wave function and only a full quantum-mechanical calculation can account for this. While such a simulation is beyond the scope of this paper, experimental evidence for the appearance of a mass transport is instead obvious: the final molecular formation process \cite{Mihajlov2012} requires that the binding energy of the formed molecule is equal to or greater than the binding energy of the Rydberg electron \cite{Boulmer1983, OKeeffe2012}. The size of the molecule is therefore about 1.7\,nm or smaller. As the measured cross section is 1000 times larger, only an efficient mass transport process can bring the two binding partners together. A second mechanism for the appearance of mass transport based on the intermediate formation of a heavy Rydberg system $Rb^++Rb^-$ with subsequent decay can not completely be ruled out. However, the poor Franck-Condon overlap between the almost flat relative wave function of the ground state atom and the rubidium core and the highly oscillatory wave function of the heavy Rydberg system (the principal quantum number which has the same binding energy as the Rydberg atom is $n'=15000$) makes this process highly unlikely. We therefore conclude that it is the contact interaction which mediates the transport. Figuratively speaking, the electron acts as a lasso that grabs a ground state atom and pulls it towards the ionic core.

An even larger fraction of the collisions lead to a decay of the Rydberg atom without the production of a molecular ion. Such processes cannot directly be tracked in our experiment, but two energetically allowed final combinations of reaction products are possible, both involving the emission of a photon: (i) the perturbation of the electron wave function by the ground state atom leads to a faster decay of the Rydberg atom. (ii) The system decays into a neutral rubidium molecule once the two partners are close enough.

The role of the electron in the molecule formation process has a close analogy to that of a catalyst in a catalytic reaction. This can be seen, looking at the Arrhenius equation
\begin{equation} 
k = n \sigma_\mathrm{coll} \bar{v}_\mathrm{th} e^{-\frac{E_\mathrm{A}}{kT}},
\label{arrhenius} 
\end{equation} 
which describes the dynamics of a chemical reaction. In reactions involving an activation barrier $E_\mathrm{A}$, the role of the catalyst is to provide a transition state that is lower in energy. The role of the electron in our reaction is not to lower the activation barrier $E_\mathrm{A}$ -- there is no such barrier -- but rather to increase the collision frequency $n \sigma_\mathrm{coll} \bar{v}_\mathrm{th}$ by altering the cross section. Moreover, the final molecular ion formation process via resonant dipole coupling  \cite{Mihajlov2012} relies on the presence of the electron. In this way, the production rate of $Rb_2^+$ molecules is increased by the presence of the Rydberg electron by orders of magnitude, just like a catalyst would do. A second condition for catalysis is that the catalyst is not consumed by the reaction. This is obviously fulfilled in our case as well. This partial catalytic process could be part of a full catalytic cycle (Fig.~\ref{fig:potential} c), where free $Rb^+$ ions and free electrons recombine into high-lying Rydberg states \cite{Vrinceanu2009} and therefore have a highly increased probability to react with surrounding ground state atoms, eventually releasing the electron again. In ultracold plasmas \cite{Killian2007} for example such catalytic cycles could take place and increase the production rate of molecular ions significantly. 

We have shown that the associative ionization in an ultracold Rydberg gas is accelerated by a factor of 1000 due to the appearance of mass transport. We attribute this transport to the attractive interaction created by scattering of the Rydberg electron from the ground state atom. The molecule formation process shows characteristics of a catalytic reaction, where the electron acts as a catalyst. Our results give a quantitative prediction for the inelastic collision rate between Rydberg atoms and ground state atoms, which constitutes an important dissipative process in ultracold Rydberg gases. At high enough densities, this process can be even larger than the spontaneous decay of the Rydberg atom. For future experiments on Rydberg dressing and studies of dissipative quantum phases of Rydberg gases beyond the frozen gas approximation, these processes should be taken into account.

We acknowledge financial support by the DFG within the SFB/TR49. O.T is funded by the graduate school of excellence MAINZ. We acknowledge discussions with Ch. Greene and T. Killian.

\bibliographystyle{apsrev4-1}




\bibliography{Molecules_Citation}

\begin{thebibliography}{25}%
\makeatletter
\providecommand \@ifxundefined [1]{%
 \@ifx{#1\undefined}
}%
\providecommand \@ifnum [1]{%
 \ifnum #1\expandafter \@firstoftwo
 \else \expandafter \@secondoftwo
 \fi
}%
\providecommand \@ifx [1]{%
 \ifx #1\expandafter \@firstoftwo
 \else \expandafter \@secondoftwo
 \fi
}%
\providecommand \natexlab [1]{#1}%
\providecommand \enquote  [1]{``#1''}%
\providecommand \bibnamefont  [1]{#1}%
\providecommand \bibfnamefont [1]{#1}%
\providecommand \citenamefont [1]{#1}%
\providecommand \href@noop [0]{\@secondoftwo}%
\providecommand \href [0]{\begingroup \@sanitize@url \@href}%
\providecommand \@href[1]{\@@startlink{#1}\@@href}%
\providecommand \@@href[1]{\endgroup#1\@@endlink}%
\providecommand \@sanitize@url [0]{\catcode `\\12\catcode `\$12\catcode
  `\&12\catcode `\#12\catcode `\^12\catcode `\_12\catcode `\%12\relax}%
\providecommand \@@startlink[1]{}%
\providecommand \@@endlink[0]{}%
\providecommand \url  [0]{\begingroup\@sanitize@url \@url }%
\providecommand \@url [1]{\endgroup\@href {#1}{\urlprefix }}%
\providecommand \urlprefix  [0]{URL }%
\providecommand \Eprint [0]{\href }%
\providecommand \doibase [0]{http://dx.doi.org/}%
\providecommand \selectlanguage [0]{\@gobble}%
\providecommand \bibinfo  [0]{\@secondoftwo}%
\providecommand \bibfield  [0]{\@secondoftwo}%
\providecommand \translation [1]{[#1]}%
\providecommand \BibitemOpen [0]{}%
\providecommand \bibitemStop [0]{}%
\providecommand \bibitemNoStop [0]{.\EOS\space}%
\providecommand \EOS [0]{\spacefactor3000\relax}%
\providecommand \BibitemShut  [1]{\csname bibitem#1\endcsname}%
\let\auto@bib@innerbib\@empty
\bibitem [{\citenamefont {Geballe}\ and\ \citenamefont
  {Oka}(1996)}]{Geballe1996}%
  \BibitemOpen
  \bibfield  {author} {\bibinfo {author} {\bibfnamefont {T.~R.}\ \bibnamefont
  {Geballe}}\ and\ \bibinfo {author} {\bibfnamefont {T.}~\bibnamefont {Oka}},\
  }\href {http://dx.doi.org/10.1038/384334a0} {\bibfield  {journal} {\bibinfo
  {journal} {Nature}\ }\textbf {\bibinfo {volume} {384}},\ \bibinfo {pages}
  {334} (\bibinfo {year} {1996})}\BibitemShut {NoStop}%
\bibitem [{\citenamefont {Dalgarno}(2005)}]{Dalgarno2005}%
  \BibitemOpen
  \bibfield  {author} {\bibinfo {author} {\bibfnamefont {A.}~\bibnamefont
  {Dalgarno}},\ }\href {http://stacks.iop.org/1742-6596/4/i=1/a=002} {\bibfield
   {journal} {\bibinfo  {journal} {Journal of Physics: Conference Series}\
  }\textbf {\bibinfo {volume} {4}},\ \bibinfo {pages} {10} (\bibinfo {year}
  {2005})}\BibitemShut {NoStop}%
\bibitem [{\citenamefont {Mihajlov}\ \emph {et~al.}(2012)\citenamefont
  {Mihajlov}, \citenamefont {Srećković}, \citenamefont {Ignjatović},\ and\
  \citenamefont {Klyucharev}}]{Mihajlov2012}%
  \BibitemOpen
  \bibfield  {author} {\bibinfo {author} {\bibfnamefont {A.}~\bibnamefont
  {Mihajlov}}, \bibinfo {author} {\bibfnamefont {V.}~\bibnamefont
  {Srećković}}, \bibinfo {author} {\bibfnamefont {L.}~\bibnamefont
  {Ignjatović}}, \ and\ \bibinfo {author} {\bibfnamefont {A.}~\bibnamefont
  {Klyucharev}},\ }\href {\doibase 10.1007/s10876-011-0438-7} {\bibfield
  {journal} {\bibinfo  {journal} {Journal of Cluster Science}\ }\textbf
  {\bibinfo {volume} {23}},\ \bibinfo {pages} {47} (\bibinfo {year}
  {2012})}\BibitemShut {NoStop}%
\bibitem [{\citenamefont {Klucharev}\ \emph {et~al.}(1980)\citenamefont
  {Klucharev}, \citenamefont {Lazarenko},\ and\ \citenamefont
  {Vujnovic}}]{Klucharev1980}%
  \BibitemOpen
  \bibfield  {author} {\bibinfo {author} {\bibfnamefont {A.~N.}\ \bibnamefont
  {Klucharev}}, \bibinfo {author} {\bibfnamefont {A.~V.}\ \bibnamefont
  {Lazarenko}}, \ and\ \bibinfo {author} {\bibfnamefont {V.}~\bibnamefont
  {Vujnovic}},\ }\href {http://stacks.iop.org/0022-3700/13/i=6/a=019}
  {\bibfield  {journal} {\bibinfo  {journal} {Journal of Physics B: Atomic and
  Molecular Physics}\ }\textbf {\bibinfo {volume} {13}},\ \bibinfo {pages}
  {1143} (\bibinfo {year} {1980})}\BibitemShut {NoStop}%
\bibitem [{\citenamefont {Cheret}\ \emph {et~al.}(1982)\citenamefont {Cheret},
  \citenamefont {Barbier}, \citenamefont {Lindinger},\ and\ \citenamefont
  {Deloche}}]{Cheret1982}%
  \BibitemOpen
  \bibfield  {author} {\bibinfo {author} {\bibfnamefont {M.}~\bibnamefont
  {Cheret}}, \bibinfo {author} {\bibfnamefont {L.}~\bibnamefont {Barbier}},
  \bibinfo {author} {\bibfnamefont {W.}~\bibnamefont {Lindinger}}, \ and\
  \bibinfo {author} {\bibfnamefont {R.}~\bibnamefont {Deloche}},\ }\href
  {http://stacks.iop.org/0022-3700/15/i=19/a=015} {\bibfield  {journal}
  {\bibinfo  {journal} {Journal of Physics B: Atomic and Molecular Physics}\
  }\textbf {\bibinfo {volume} {15}},\ \bibinfo {pages} {3463} (\bibinfo {year}
  {1982})}\BibitemShut {NoStop}%
\bibitem [{\citenamefont {Barbier}\ and\ \citenamefont
  {Cheret}(1987)}]{Barbier1987}%
  \BibitemOpen
  \bibfield  {author} {\bibinfo {author} {\bibfnamefont {L.}~\bibnamefont
  {Barbier}}\ and\ \bibinfo {author} {\bibfnamefont {M.}~\bibnamefont
  {Cheret}},\ }\href {http://stacks.iop.org/0022-3700/20/i=6/a=011} {\bibfield
  {journal} {\bibinfo  {journal} {Journal of Physics B: Atomic and Molecular
  Physics}\ }\textbf {\bibinfo {volume} {20}},\ \bibinfo {pages} {1229}
  (\bibinfo {year} {1987})}\BibitemShut {NoStop}%
\bibitem [{\citenamefont {Fermi}(1934)}]{Fermi1934}%
  \BibitemOpen
  \bibfield  {author} {\bibinfo {author} {\bibfnamefont {E.}~\bibnamefont
  {Fermi}},\ }\href@noop {} {\bibfield  {journal} {\bibinfo  {journal} {Nuovo
  Cimento}\ }\textbf {\bibinfo {volume} {11}},\ \bibinfo {pages} {157}
  (\bibinfo {year} {1934})}\BibitemShut {NoStop}%
\bibitem [{\citenamefont {Bendkowsky}\ \emph {et~al.}(2009)\citenamefont
  {Bendkowsky}, \citenamefont {Butscher}, \citenamefont {Nipper}, \citenamefont
  {Shaffer}, \citenamefont {L\"ow},\ and\ \citenamefont
  {Pfau}}]{Bendkowsky2009}%
  \BibitemOpen
  \bibfield  {author} {\bibinfo {author} {\bibfnamefont {V.}~\bibnamefont
  {Bendkowsky}}, \bibinfo {author} {\bibfnamefont {B.}~\bibnamefont
  {Butscher}}, \bibinfo {author} {\bibfnamefont {J.}~\bibnamefont {Nipper}},
  \bibinfo {author} {\bibfnamefont {J.~P.}\ \bibnamefont {Shaffer}}, \bibinfo
  {author} {\bibfnamefont {R.}~\bibnamefont {L\"ow}}, \ and\ \bibinfo {author}
  {\bibfnamefont {T.}~\bibnamefont {Pfau}},\ }\href
  {http://dx.doi.org/10.1038/nature07945} {\bibfield  {journal} {\bibinfo
  {journal} {Nature}\ }\textbf {\bibinfo {volume} {458}},\ \bibinfo {pages}
  {1005} (\bibinfo {year} {2009})}\BibitemShut {NoStop}%
\bibitem [{\citenamefont {Booth}\ \emph {et~al.}(2014)\citenamefont {Booth},
  \citenamefont {Rittenhouse}, \citenamefont {Yang}, \citenamefont
  {Sadeghpour},\ and\ \citenamefont {Shaffer}}]{Booth2014}%
  \BibitemOpen
  \bibfield  {author} {\bibinfo {author} {\bibfnamefont {D.}~\bibnamefont
  {Booth}}, \bibinfo {author} {\bibfnamefont {S.}~\bibnamefont {Rittenhouse}},
  \bibinfo {author} {\bibfnamefont {J.}~\bibnamefont {Yang}}, \bibinfo {author}
  {\bibfnamefont {H.}~\bibnamefont {Sadeghpour}}, \ and\ \bibinfo {author}
  {\bibfnamefont {J.}~\bibnamefont {Shaffer}},\ }\href@noop {} {\  (\bibinfo
  {year} {2014})},\ \Eprint {http://arxiv.org/abs/arXiv:1411.5291}
  {arXiv:1411.5291} \BibitemShut {NoStop}%
\bibitem [{\citenamefont {Balewski}\ \emph {et~al.}(2013)\citenamefont
  {Balewski}, \citenamefont {Krupp}, \citenamefont {Gaj}, \citenamefont
  {Peter}, \citenamefont {Buchler}, \citenamefont {Low}, \citenamefont
  {Hofferberth},\ and\ \citenamefont {Pfau}}]{Balewski2013}%
  \BibitemOpen
  \bibfield  {author} {\bibinfo {author} {\bibfnamefont {J.~B.}\ \bibnamefont
  {Balewski}}, \bibinfo {author} {\bibfnamefont {A.~T.}\ \bibnamefont {Krupp}},
  \bibinfo {author} {\bibfnamefont {A.}~\bibnamefont {Gaj}}, \bibinfo {author}
  {\bibfnamefont {D.}~\bibnamefont {Peter}}, \bibinfo {author} {\bibfnamefont
  {H.~P.}\ \bibnamefont {Buchler}}, \bibinfo {author} {\bibfnamefont
  {R.}~\bibnamefont {Low}}, \bibinfo {author} {\bibfnamefont {S.}~\bibnamefont
  {Hofferberth}}, \ and\ \bibinfo {author} {\bibfnamefont {T.}~\bibnamefont
  {Pfau}},\ }\href {http://dx.doi.org/10.1038/nature12592} {\bibfield
  {journal} {\bibinfo  {journal} {Nature}\ }\textbf {\bibinfo {volume} {502}},\
  \bibinfo {pages} {664} (\bibinfo {year} {2013})}\BibitemShut {NoStop}%
\bibitem [{\citenamefont {Schau{\ss}}\ \emph {et~al.}(2012)\citenamefont
  {Schau{\ss}}, \citenamefont {Cheneau}, \citenamefont {Endres}, \citenamefont
  {Fukuhara}, \citenamefont {Hild}, \citenamefont {Omran}, \citenamefont
  {Pohl}, \citenamefont {Gross}, \citenamefont {Kuhr},\ and\ \citenamefont
  {Bloch}}]{Schausz2012}%
  \BibitemOpen
  \bibfield  {author} {\bibinfo {author} {\bibfnamefont {P.}~\bibnamefont
  {Schau{\ss}}}, \bibinfo {author} {\bibfnamefont {M.}~\bibnamefont {Cheneau}},
  \bibinfo {author} {\bibfnamefont {M.}~\bibnamefont {Endres}}, \bibinfo
  {author} {\bibfnamefont {T.}~\bibnamefont {Fukuhara}}, \bibinfo {author}
  {\bibfnamefont {S.}~\bibnamefont {Hild}}, \bibinfo {author} {\bibfnamefont
  {A.}~\bibnamefont {Omran}}, \bibinfo {author} {\bibfnamefont
  {T.}~\bibnamefont {Pohl}}, \bibinfo {author} {\bibfnamefont {C.}~\bibnamefont
  {Gross}}, \bibinfo {author} {\bibfnamefont {S.}~\bibnamefont {Kuhr}}, \ and\
  \bibinfo {author} {\bibfnamefont {I.}~\bibnamefont {Bloch}},\ }\href
  {http://dx.doi.org/10.1038/nature11596} {\bibfield  {journal} {\bibinfo
  {journal} {Nature}\ }\textbf {\bibinfo {volume} {491}},\ \bibinfo {pages}
  {87} (\bibinfo {year} {2012})}\BibitemShut {NoStop}%
\bibitem [{\citenamefont {Schempp}\ \emph {et~al.}(2014)\citenamefont
  {Schempp}, \citenamefont {G\"unter}, \citenamefont {Robert-de Saint-Vincent},
  \citenamefont {Hofmann}, \citenamefont {Breyel}, \citenamefont {Komnik},
  \citenamefont {Sch\"onleber}, \citenamefont {G\"arttner}, \citenamefont
  {Evers}, \citenamefont {Whitlock},\ and\ \citenamefont
  {Weidem\"uller}}]{Schempp2014}%
  \BibitemOpen
  \bibfield  {author} {\bibinfo {author} {\bibfnamefont {H.}~\bibnamefont
  {Schempp}}, \bibinfo {author} {\bibfnamefont {G.}~\bibnamefont {G\"unter}},
  \bibinfo {author} {\bibfnamefont {M.}~\bibnamefont {Robert-de
  Saint-Vincent}}, \bibinfo {author} {\bibfnamefont {C.~S.}\ \bibnamefont
  {Hofmann}}, \bibinfo {author} {\bibfnamefont {D.}~\bibnamefont {Breyel}},
  \bibinfo {author} {\bibfnamefont {A.}~\bibnamefont {Komnik}}, \bibinfo
  {author} {\bibfnamefont {D.~W.}\ \bibnamefont {Sch\"onleber}}, \bibinfo
  {author} {\bibfnamefont {M.}~\bibnamefont {G\"arttner}}, \bibinfo {author}
  {\bibfnamefont {J.}~\bibnamefont {Evers}}, \bibinfo {author} {\bibfnamefont
  {S.}~\bibnamefont {Whitlock}}, \ and\ \bibinfo {author} {\bibfnamefont
  {M.}~\bibnamefont {Weidem\"uller}},\ }\href {\doibase
  10.1103/PhysRevLett.112.013002} {\bibfield  {journal} {\bibinfo  {journal}
  {Phys. Rev. Lett.}\ }\textbf {\bibinfo {volume} {112}},\ \bibinfo {pages}
  {013002} (\bibinfo {year} {2014})}\BibitemShut {NoStop}%
\bibitem [{\citenamefont {Weber}\ \emph {et~al.}(2015)\citenamefont {Weber},
  \citenamefont {H\"oning}, \citenamefont {Niederpr\"um}, \citenamefont
  {Manthey}, \citenamefont {Thomas}, \citenamefont {Guarrera}, \citenamefont
  {Fleischhauer}, \citenamefont {Barontini},\ and\ \citenamefont
  {Ott}}]{Weber2015}%
  \BibitemOpen
  \bibfield  {author} {\bibinfo {author} {\bibfnamefont {T.~M.}\ \bibnamefont
  {Weber}}, \bibinfo {author} {\bibfnamefont {M.}~\bibnamefont {H\"oning}},
  \bibinfo {author} {\bibfnamefont {T.}~\bibnamefont {Niederpr\"um}}, \bibinfo
  {author} {\bibfnamefont {T.}~\bibnamefont {Manthey}}, \bibinfo {author}
  {\bibfnamefont {O.}~\bibnamefont {Thomas}}, \bibinfo {author} {\bibfnamefont
  {V.}~\bibnamefont {Guarrera}}, \bibinfo {author} {\bibfnamefont
  {M.}~\bibnamefont {Fleischhauer}}, \bibinfo {author} {\bibfnamefont
  {G.}~\bibnamefont {Barontini}}, \ and\ \bibinfo {author} {\bibfnamefont
  {H.}~\bibnamefont {Ott}},\ }\href {http://dx.doi.org/10.1038/nphys3214}
  {\bibfield  {journal} {\bibinfo  {journal} {Nat Phys}\ }\textbf {\bibinfo
  {volume} {11}},\ \bibinfo {pages} {157} (\bibinfo {year} {2015})}\BibitemShut
  {NoStop}%
\bibitem [{\citenamefont {Studer}\ and\ \citenamefont
  {Curran}(2014)}]{Studer2014}%
  \BibitemOpen
  \bibfield  {author} {\bibinfo {author} {\bibfnamefont {A.}~\bibnamefont
  {Studer}}\ and\ \bibinfo {author} {\bibfnamefont {D.~P.}\ \bibnamefont
  {Curran}},\ }\href {http://dx.doi.org/10.1038/nchem.2031} {\bibfield
  {journal} {\bibinfo  {journal} {Nat Chem}\ }\textbf {\bibinfo {volume} {6}},\
  \bibinfo {pages} {765} (\bibinfo {year} {2014})}\BibitemShut {NoStop}%
\bibitem [{\citenamefont {Manthey}\ \emph {et~al.}(2014)\citenamefont
  {Manthey}, \citenamefont {Weber}, \citenamefont {Niederpr\"um}, \citenamefont
  {Langer}, \citenamefont {Guarrera}, \citenamefont {Barontini},\ and\
  \citenamefont {Ott}}]{Manthey2014}%
  \BibitemOpen
  \bibfield  {author} {\bibinfo {author} {\bibfnamefont {T.}~\bibnamefont
  {Manthey}}, \bibinfo {author} {\bibfnamefont {T.~M.}\ \bibnamefont {Weber}},
  \bibinfo {author} {\bibfnamefont {T.}~\bibnamefont {Niederpr\"um}}, \bibinfo
  {author} {\bibfnamefont {P.}~\bibnamefont {Langer}}, \bibinfo {author}
  {\bibfnamefont {V.}~\bibnamefont {Guarrera}}, \bibinfo {author}
  {\bibfnamefont {G.}~\bibnamefont {Barontini}}, \ and\ \bibinfo {author}
  {\bibfnamefont {H.}~\bibnamefont {Ott}},\ }\href
  {http://stacks.iop.org/1367-2630/16/i=8/a=083034} {\bibfield  {journal}
  {\bibinfo  {journal} {New Journal of Physics}\ }\textbf {\bibinfo {volume}
  {16}},\ \bibinfo {pages} {083034} (\bibinfo {year} {2014})}\BibitemShut
  {NoStop}%
\bibitem [{\citenamefont {Beterov}\ \emph {et~al.}(2007)\citenamefont
  {Beterov}, \citenamefont {Tretyakov}, \citenamefont {Ryabtsev}, \citenamefont
  {Ekers},\ and\ \citenamefont {Bezuglov}}]{Beterov2007}%
  \BibitemOpen
  \bibfield  {author} {\bibinfo {author} {\bibfnamefont {I.~I.}\ \bibnamefont
  {Beterov}}, \bibinfo {author} {\bibfnamefont {D.~B.}\ \bibnamefont
  {Tretyakov}}, \bibinfo {author} {\bibfnamefont {I.~I.}\ \bibnamefont
  {Ryabtsev}}, \bibinfo {author} {\bibfnamefont {A.}~\bibnamefont {Ekers}}, \
  and\ \bibinfo {author} {\bibfnamefont {N.~N.}\ \bibnamefont {Bezuglov}},\
  }\href {\doibase 10.1103/PhysRevA.75.052720} {\bibfield  {journal} {\bibinfo
  {journal} {Phys. Rev. A}\ }\textbf {\bibinfo {volume} {75}},\ \bibinfo
  {pages} {052720} (\bibinfo {year} {2007})}\BibitemShut {NoStop}%
\bibitem [{\citenamefont {Urban}\ \emph {et~al.}(2009)\citenamefont {Urban},
  \citenamefont {Johnson}, \citenamefont {Henage}, \citenamefont {Isenhower},
  \citenamefont {Yavuz}, \citenamefont {Walker},\ and\ \citenamefont
  {Saffman}}]{Urban2009}%
  \BibitemOpen
  \bibfield  {author} {\bibinfo {author} {\bibfnamefont {E.}~\bibnamefont
  {Urban}}, \bibinfo {author} {\bibfnamefont {T.~A.}\ \bibnamefont {Johnson}},
  \bibinfo {author} {\bibfnamefont {T.}~\bibnamefont {Henage}}, \bibinfo
  {author} {\bibfnamefont {L.}~\bibnamefont {Isenhower}}, \bibinfo {author}
  {\bibfnamefont {D.~D.}\ \bibnamefont {Yavuz}}, \bibinfo {author}
  {\bibfnamefont {T.~G.}\ \bibnamefont {Walker}}, \ and\ \bibinfo {author}
  {\bibfnamefont {M.}~\bibnamefont {Saffman}},\ }\href
  {http://dx.doi.org/10.1038/nphys1178} {\bibfield  {journal} {\bibinfo
  {journal} {Nat Phys}\ }\textbf {\bibinfo {volume} {5}},\ \bibinfo {pages}
  {110} (\bibinfo {year} {2009})}\BibitemShut {NoStop}%
\bibitem [{\citenamefont {Gaetan}\ \emph {et~al.}(2009)\citenamefont {Gaetan},
  \citenamefont {Miroshnychenko}, \citenamefont {Wilk}, \citenamefont {Chotia},
  \citenamefont {Viteau}, \citenamefont {Comparat}, \citenamefont {Pillet},
  \citenamefont {Browaeys},\ and\ \citenamefont {Grangier}}]{Gaetan2009}%
  \BibitemOpen
  \bibfield  {author} {\bibinfo {author} {\bibfnamefont {A.}~\bibnamefont
  {Gaetan}}, \bibinfo {author} {\bibfnamefont {Y.}~\bibnamefont
  {Miroshnychenko}}, \bibinfo {author} {\bibfnamefont {T.}~\bibnamefont
  {Wilk}}, \bibinfo {author} {\bibfnamefont {A.}~\bibnamefont {Chotia}},
  \bibinfo {author} {\bibfnamefont {M.}~\bibnamefont {Viteau}}, \bibinfo
  {author} {\bibfnamefont {D.}~\bibnamefont {Comparat}}, \bibinfo {author}
  {\bibfnamefont {P.}~\bibnamefont {Pillet}}, \bibinfo {author} {\bibfnamefont
  {A.}~\bibnamefont {Browaeys}}, \ and\ \bibinfo {author} {\bibfnamefont
  {P.}~\bibnamefont {Grangier}},\ }\href {http://dx.doi.org/10.1038/nphys1183}
  {\bibfield  {journal} {\bibinfo  {journal} {Nat Phys}\ }\textbf {\bibinfo
  {volume} {5}},\ \bibinfo {pages} {115} (\bibinfo {year} {2009})}\BibitemShut
  {NoStop}%
\bibitem [{Note1()}]{Note1}%
  \BibitemOpen
  \bibinfo {note} {The Rydberg blockade becomes relevant when the peak density
  of the cloud $n_0$ exceeds the volume of the blockade sphere $r_\protect
  \mathrm {B}^3$}\BibitemShut {NoStop}%
\bibitem [{\citenamefont {Greene}\ \emph {et~al.}(2000)\citenamefont {Greene},
  \citenamefont {Dickinson},\ and\ \citenamefont {Sadeghpour}}]{Greene2000}%
  \BibitemOpen
  \bibfield  {author} {\bibinfo {author} {\bibfnamefont {C.~H.}\ \bibnamefont
  {Greene}}, \bibinfo {author} {\bibfnamefont {A.~S.}\ \bibnamefont
  {Dickinson}}, \ and\ \bibinfo {author} {\bibfnamefont {H.~R.}\ \bibnamefont
  {Sadeghpour}},\ }\href {\doibase 10.1103/PhysRevLett.85.2458} {\bibfield
  {journal} {\bibinfo  {journal} {Phys. Rev. Lett.}\ }\textbf {\bibinfo
  {volume} {85}},\ \bibinfo {pages} {2458} (\bibinfo {year}
  {2000})}\BibitemShut {NoStop}%
\bibitem [{\citenamefont {Anderson}\ \emph {et~al.}(2014)\citenamefont
  {Anderson}, \citenamefont {Miller},\ and\ \citenamefont
  {Raithel}}]{Anderson2014}%
  \BibitemOpen
  \bibfield  {author} {\bibinfo {author} {\bibfnamefont {D.~A.}\ \bibnamefont
  {Anderson}}, \bibinfo {author} {\bibfnamefont {S.~A.}\ \bibnamefont
  {Miller}}, \ and\ \bibinfo {author} {\bibfnamefont {G.}~\bibnamefont
  {Raithel}},\ }\href@noop {} {\  (\bibinfo {year} {2014})},\ \Eprint
  {http://arxiv.org/abs/1409.2543} {1409.2543} \BibitemShut {NoStop}%
\bibitem [{\citenamefont {Boulmer}\ \emph {et~al.}(1983)\citenamefont
  {Boulmer}, \citenamefont {Bonanno},\ and\ \citenamefont
  {Weiner}}]{Boulmer1983}%
  \BibitemOpen
  \bibfield  {author} {\bibinfo {author} {\bibfnamefont {J.}~\bibnamefont
  {Boulmer}}, \bibinfo {author} {\bibfnamefont {R.}~\bibnamefont {Bonanno}}, \
  and\ \bibinfo {author} {\bibfnamefont {J.}~\bibnamefont {Weiner}},\ }\href
  {http://stacks.iop.org/0022-3700/16/i=16/a=021} {\bibfield  {journal}
  {\bibinfo  {journal} {Journal of Physics B: Atomic and Molecular Physics}\
  }\textbf {\bibinfo {volume} {16}},\ \bibinfo {pages} {3015} (\bibinfo {year}
  {1983})}\BibitemShut {NoStop}%
\bibitem [{\citenamefont {O'Keeffe}\ \emph {et~al.}(2012)\citenamefont
  {O'Keeffe}, \citenamefont {Bolognesi}, \citenamefont {Avaldi}, \citenamefont
  {Moise}, \citenamefont {Richter}, \citenamefont {Mihajlov}, \citenamefont
  {Sre\ifmmode \acute{c}\else \'{c}\fi{}kovi\ifmmode~\acute{c}\else
  \'{c}\fi{}},\ and\ \citenamefont {Ignjatovi\ifmmode~\acute{c}\else
  \'{c}\fi{}}}]{OKeeffe2012}%
  \BibitemOpen
  \bibfield  {author} {\bibinfo {author} {\bibfnamefont {P.}~\bibnamefont
  {O'Keeffe}}, \bibinfo {author} {\bibfnamefont {P.}~\bibnamefont {Bolognesi}},
  \bibinfo {author} {\bibfnamefont {L.}~\bibnamefont {Avaldi}}, \bibinfo
  {author} {\bibfnamefont {A.}~\bibnamefont {Moise}}, \bibinfo {author}
  {\bibfnamefont {R.}~\bibnamefont {Richter}}, \bibinfo {author} {\bibfnamefont
  {A.~A.}\ \bibnamefont {Mihajlov}}, \bibinfo {author} {\bibfnamefont {V.~A.}\
  \bibnamefont {Sre\ifmmode \acute{c}\else
  \'{c}\fi{}kovi\ifmmode~\acute{c}\else \'{c}\fi{}}}, \ and\ \bibinfo {author}
  {\bibfnamefont {L.~M.}\ \bibnamefont {Ignjatovi\ifmmode~\acute{c}\else
  \'{c}\fi{}}},\ }\href {\doibase 10.1103/PhysRevA.85.052705} {\bibfield
  {journal} {\bibinfo  {journal} {Phys. Rev. A}\ }\textbf {\bibinfo {volume}
  {85}},\ \bibinfo {pages} {052705} (\bibinfo {year} {2012})}\BibitemShut
  {NoStop}%
\bibitem [{\citenamefont {Vrinceanu}\ \emph {et~al.}(2009)\citenamefont
  {Vrinceanu}, \citenamefont {Sadeghpour},\ and\ \citenamefont
  {Pohl}}]{Vrinceanu2009}%
  \BibitemOpen
  \bibfield  {author} {\bibinfo {author} {\bibfnamefont {D.}~\bibnamefont
  {Vrinceanu}}, \bibinfo {author} {\bibfnamefont {H.~R.}\ \bibnamefont
  {Sadeghpour}}, \ and\ \bibinfo {author} {\bibfnamefont {T.}~\bibnamefont
  {Pohl}},\ }\href {http://stacks.iop.org/1742-6596/194/i=1/a=012067}
  {\bibfield  {journal} {\bibinfo  {journal} {Journal of Physics: Conference
  Series}\ }\textbf {\bibinfo {volume} {194}},\ \bibinfo {pages} {012067}
  (\bibinfo {year} {2009})}\BibitemShut {NoStop}%
\bibitem [{\citenamefont {Killian}(2007)}]{Killian2007}%
  \BibitemOpen
  \bibfield  {author} {\bibinfo {author} {\bibfnamefont {T.~C.}\ \bibnamefont
  {Killian}},\ }\href {\doibase 10.1126/science.1130556} {\bibfield  {journal}
  {\bibinfo  {journal} {Science}\ }\textbf {\bibinfo {volume} {316}},\ \bibinfo
  {pages} {705} (\bibinfo {year} {2007})}\BibitemShut {NoStop}%
\end{thebibliography}%

\end{document}